\documentclass{jetpl}

\usepackage{graphicx}

\twocolumn
\lat

\begin{document}
\title{Electric field dependence of thermal conductivity of
a granular superconductor: Giant field-induced effects predicted}
 \rtitle{Thermal conductivity of a granular superconductor}
 \sodtitle{Thermal conductivity of a granular superconductor}

\author{S.A. Sergeenkov}

\address{Bogoliubov Laboratory of Theoretical Physics,
Joint Institute for Nuclear Research, 141980 Dubna, Moscow Region,
Russia}

\dates{1 July 2002}{*}

\abstract {The temperature and electric field dependence of
electronic contribution to the thermal conductivity (TC) of a
granular superconductor is considered within a 3D model of inductive
Josephson junction arrays. In addition to a low-temperature maximum
of zero-field TC $\kappa (T,0)$ (controlled by mutual inductance
$L_0$ and normal state resistivity $R_n$), the model predicts two
major effects in applied electric field: (i) {\it decrease} of the
{\it linear} TC, and (ii) giant {\it enhancement} of the {\it
nonlinear} (i.e. $\nabla T$ - dependent) TC with $\Delta \kappa
(T,E)/\kappa (T,0)$ reaching $500\%$ for parallel electric fields
$E\simeq E_T$ ($E_T=S_0|\nabla T|$ is an "intrinsic" thermoelectric
field). A possiblity of experimental observation of the predicted
effects in granular superconductors is discussed.}


\maketitle

{\bf 1. Introduction.} Inspired by new possibilities offered by the
cutting-edge nanotechnologies, the experimental and theoretical
physics of increasingly sophisticated mesoscopic quantum devices
(heavily based on Josephson junctions and their arrays) is becoming
one of the most exciting and rapidly growing areas of modern science
~\cite{1,2,3}. In addition to the traditional fields of expertise
(such as granular superconductors~\cite{2}), Josephson junction
arrays (JJAs) are actively used for testing principally novel ideas
(like, e.g., topologically protected quantum bits~\cite{3}) in a bid
to solve probably one of the most challenging problems in quantum
computing. Though traditionally, the main emphasis in studying JJAs
has been on their behavior in applied magnetic fields, since recently
a special attention has been given to the so-called electric field
effects (FEs) in JJs and granular superconductors~\cite{4,5,6,7,8,9}.
The unusually strong FEs observed in bulk high-$T_c$ superconducting
(HTS) ceramics~\cite{4} (including a substantial enhancement of the
critical current, reaching $\Delta I_c(E)/I_c(0)=100\%$ for
$E=10^7V/m$) have been attributed to a crucial modification of the
original weak-links structure under the influence of very strong
electric fields. This hypothesis has been corroborated by further
investigations, both experimental (through observation of the
correlation between the critical current behavior and type of weak
links~\cite{5}) and theoretical (by studying the FEs in $SNS$-type
structures~\cite{6} and $d$-wave granular superconductors~\cite{7}).
Among other interesting field induced effects, one can mention the
 FE-based Josephson transistor~\cite{8} and
Josephson analog of the {\it magnetoelectric effect}~\cite{9}
(electric field generation of Josephson magnetic moment in zero
magnetic field). At the same time, very little is known about
influence of electric fields on thermal transport properties of
granular superconductors. In an attempt to shed some light on this
intresting and important (for potential applications) problem, in
this Letter we present a theoretical study of the electric field and
temperature dependence of electronic contribution to thermal
conductivity (TC) $\kappa$ of a granular superconductor (described by
a 3D model of inductive JJAs). As we shall see below, in addition to
a low-temperature maximum of zero-field TC $\kappa (T,0)$ (controlled
by the mutual inductance $L_0$ and normal state resistivity $R_n$),
the model predicts unusually strong (giant) field-induced effects in
the behavior of {\it nonlinear} (i.e. $\nabla T$ - dependent) TC. In
particular, the absolute values of the TC enhancement $\Delta \kappa
(T,E)/\kappa (T,0)$ are estimated to reach up to $500\%$ for
relatively low (in comparison with the fields needed to observe a
critical current enhancement~\cite{4,5}) applied electric fields $E$
matching an intrinsic thermoelectric field $E_T=S_0|\nabla T|$. The
estimates of the model parameters suggest quite an optimistic
possibility to observe the predicted effects in granular
superconductors and JJAs.

{\bf 2. The model.} To adequately describe a thermodynamic behavior
of a real granular superconductor for all temperatures and under a
simultaneous influence of arbitrary electric field ${\bf E}$ and
thermal gradient $\nabla T$, we consider one of the numerous versions
of the 3D JJAs models based on the following Hamiltonian
\begin{equation}
{\cal H}(t)={\cal H}_T(t)+{\cal H}_L(t)+{\cal H}_E(t),
\end{equation}
where
\begin{equation}
{\cal H}_T(t)=\sum_{ij}^NJ_{ij}[1-\cos \phi _{ij}(t)]
\end{equation}
is the well-known tunneling Hamiltonian,
\begin{equation}
{\cal H}_L(t)=\sum_{ij}^N\frac{\Phi _{ij}^2(t)}{2L_{ij}}
\end{equation}
accounts for a mutual inductance $L_{ij}$ between grains (and
controls the normal state value of the thermal conductivity, see
below) with $\Phi _{ij}(t)=(\hbar /2e)\phi _{ij}(t)$ being the total
magnetic flux through an array, and finally
\begin{equation}
{\cal H}_E(t)=-{\bf p}(t){\bf E}
\end{equation}
describes electric field induced polarization contribution, where the
polarization operator
\begin{equation}
{\bf p}(t)=-2e\sum_{i=1}^Nn_i(t){\bf r}_i
\end{equation}
Here $n_i$ is the pair number operator, and ${\bf r}_i$ is the
coordinate of the center of the grain.

As usual, the tunneling Hamiltonian ${\cal H}_T(t)$ describes a
short-range interaction between $N$ superconducting grains, arranged
in a 3D lattice with coordinates ${\bf r}_i=(x_i,y_i,z_i)$. The
grains are separated by insulating boundaries producing temperature
dependent Josephson coupling $J_{ij}(T)=J_{ij}(0)F(T)$ with
\begin{equation}
F(T)=\frac{\Delta (T)}{\Delta (0)}\tanh \left [\frac{\Delta
(T)}{2k_BT}\right ]
\end{equation}
and $J_{ij}(0)=[\Delta (0)/2](R_0/R_{ij})$ where $\Delta (T)$ is the
temperature dependent gap parameter, $R_0=h/4e^2$ is the quantum
resistance, and $R_{ij}$ is the resistance between grains in their
normal state assumed~\cite{10} to vary exponentially with the
distance ${\bf r}_{ij}$ between neighboring grains, i.e.
$R_{ij}^{-1}=R_n^{-1}\exp(-r_{ij}/d)$ (where $d$ is of the order of
an average grain size).

As is well-known~\cite{2,10}, a constant electric field ${\bf E}$ and
a thermal gradient $\nabla T$ applied to a JJA cause a time evolution
of the initial phase difference $\phi _{ij}^0=\phi _i-\phi _j$ as
follows
\begin{equation}
\phi _{ij}(t)=\phi _{ij}^0+\omega _{ij}({\bf E},\nabla T)t
\end{equation}
Here $\omega _{ij}=2e({\bf E}-{\bf E}_T){\bf r}_{ij}/\hbar$ where
${\bf E}_T=S_0\nabla T$ is an "intrinsic" thermoelectric field with
$S_0$ being a zero-field value of the Seebeck coefficient.

{\bf 3. Linear thermal conductivity (Fourier law).} We start our
consideration by discussing the temperature behavior of the
conventional (that is {\it linear}) thermal conductivity of a
granular superconductor in arbitrary applied electric field ${\bf E}$
paying a special attention to its evolution with a mutual inductance
$L_{ij}$. For simplicity, in what follows we limit our consideration
to the longitudinal component of the total thermal flux ${\bf Q}(t)$
which is defined (in a q-space representation) via the total energy
conservation law as follows
\begin{equation}
{\bf Q}(t)\equiv \lim_{{\bf q} \to 0} \left [i\frac{{\bf q}}{{\bf
q}^2}{\dot{\cal H}_{\bf q}}(t)\right ],
\end{equation}
where ${\dot{\cal H}_{\bf q}}=\partial {\cal H}_{\bf q}/\partial t$
with
\begin{equation}
{\cal H}_{\bf q}(t)=\frac{1}{v}\int d^3x e^{i{\bf q}{\bf r}}{\cal
H}({\bf r},t)
\end{equation}
Here $v=8\pi d^3$ is properly defined normalization volume, and we
made a usual substitution $\frac{1}{N}\sum_{ij}A(r_{ij},t) \to
\frac{1}{v}\int d^3x A({\bf r},t)$ valid in the long-wavelength
approximation (${\bf q} \to 0$).

In turn, the above-introduced heat flux ${\bf Q}(t)$ is related to
the appropriate components of the {\it linear} thermal conductivity
(LTC) tensor $\kappa _{\alpha \beta}$ as follows (hereafter,
$\{\alpha ,\beta \}=x,y,z$)
\begin{equation}
\kappa _{\alpha \beta}(T,{\bf E})\equiv -\frac{1}{V}\left
[\frac{\partial \overline{<Q_{\alpha}>}}{\partial (\nabla
_{\beta}T)}\right ]_{\nabla T=0},
\end{equation}
where
\begin{equation}
\overline{<Q_{\alpha}>}=\frac{1}{\tau}\int_0^\tau dt<Q_{\alpha}(t)>
\end{equation}
Here $V$ is a sample's volume, $\tau$ is a characteristic Josephson
time for the network, and $<...>$ denotes the thermodynamic averaging
over the initial phase differences $\phi _{ij}^0$
\begin{equation}
<A(\phi _{ij}^0)>=\frac{1}{Z}\int_0^{\pi}\prod _{ij}
d\phi _{ij}^0 A(\phi _{ij}^0)e^{-\beta H_0}
\end{equation}
with an effective Hamiltonian
\begin{equation}
H_0[\phi _{ij}^0]=\int_0^{\tau}\frac{dt}{\tau}\int
\frac{d^3x}{v}{\cal H}({\bf r},t)
\end{equation}
Here, $\beta =1/k_BT$, and $Z=\int_0^{\pi}\prod _{ij}d\phi _{ij}^0
e^{-\beta H_0}$ is the partition function. The above-defined
averaging procedure allows us to study the temperature evolution of
the system.

Taking into account that in JJAs~\cite{11} $L_{ij}\propto R_{ij}$, we
obtain $L_{ij}=L_{0}\exp(r_{ij}/d)$ for the explicit $r$-dependence
of the weak-link inductance in our model. Finally, in view of
Eqs.(1)-(13), and making use of the usual "phase-number" commutation
relation, $[\phi _i,n_j]=i\delta _{ij}$, we find the following
analytical expression for the temperature and electric field
dependence of the electronic contribution to {\it linear} thermal
conductivity of a granular superconductor
\begin{equation}
\kappa _{\alpha \beta}(T,{\bf E})=\kappa _0[\delta _{\alpha
\beta}\eta (T,\epsilon )+\beta _L(T)\nu (T,\epsilon )f_{\alpha
\beta}(\epsilon )]
\end{equation}
where
\begin{equation}
f_{\alpha \beta}(\epsilon )=\frac{1}{4}\left [\delta _{\alpha
\beta}A(\epsilon)-\epsilon _{\alpha}\epsilon
_{\beta}B(\epsilon)\right ]
\end{equation}
with
\begin{equation}
A(\epsilon)=\frac{5+3\epsilon ^2}{(1+\epsilon
^2)^2}+\frac{3}{\epsilon}\tan ^{-1}\epsilon
\end{equation}
and
\begin{equation}
B(\epsilon)=\frac{3\epsilon ^4+8\epsilon ^2-3}{\epsilon ^2(1+\epsilon
^2)^3}+\frac{3}{\epsilon ^3}\tan ^{-1}\epsilon
\end{equation}
Here, $\kappa _0=Nd^2S_0\Phi _0/VL_{0}$, $\beta _L(T)=2\pi I_c(T)
L_{0}/\Phi _0$ with $I_c(T)=(2e/\hbar )J(T)$ being the critical
current (we neglect a possible field dependence of $I_c$ because, as
we shall see below, the characteristic fields where thermal
conductivity exhibits most interesting behavior are much lower than
those needed to produce a tangible change of the critical
current~\cite{4}); $\epsilon \equiv \sqrt{\epsilon _x^2+\epsilon
_y^2+\epsilon _z^2}$ with $\epsilon _{\alpha} =E_{\alpha}/E_0$, and
$E_0=\hbar /(2ed\tau )$ is a characteristic electric field. In turn,
the above-introduced "order parameters" of the system, $\eta
(T,\epsilon )\equiv <\phi _{ij}^0>$ and $\nu (T,\epsilon )\equiv
<\sin \phi _{ij}^0>$, are defined as follows
\begin{equation}
\eta (T, \epsilon
)=\frac{\pi}{2}-\frac{4}{\pi}\sum_{n=0}^{\infty}\frac{1}{(2n+1)^2}
\left [\frac{I_{2n+1}(\beta _E)}{I_0(\beta _E)}\right ]
\end{equation}
and
\begin{equation}
\nu (T, \epsilon )= \frac{\sinh \beta _E}{\beta _EI_0(\beta _E)},
\end{equation}
where
\begin{equation}
\beta _E(T,\epsilon )=\frac{\beta J(T)}{2}\left (\frac{1}{1+\epsilon
^2}+\frac{1}{\epsilon}\tan ^{-1}\epsilon \right)
\end{equation}
Here $J(T)=J(0)F(T)$ with $J(0)=(\Delta _0/2)(R_0/R_n)$ and $F(T)$
given by Eq.(6); $I_n(x)$ stand for the appropriate modified Bessel
functions.

{\it 3.1. Zero-field effects.} Turning to the discussion of the
obtained results, we start with a more simple zero-field case. The
relevant parameters affecting the behavior of the LTC in this
particular case include the mutual inductance $L_{0}$ and the normal
state resistance between grains $R_n$. For the temperature dependence
of the Josephson energy (see Eq.(6)), we used the
well-known~\cite{12} approximation for the BCS gap parameter, valid
for all temperatures, $\Delta (T)=\Delta (0)\tanh \left (\gamma
\sqrt{\frac{T_c-T}{T}}\right )$ with $\gamma =2.2$.

\begin{figure}
 \centerline{\includegraphics[width=90mm]{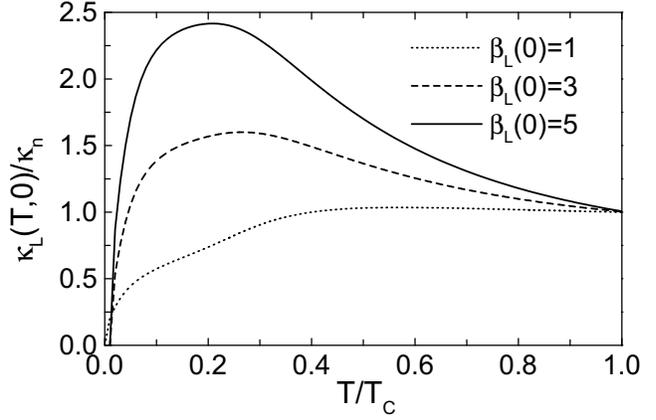}}
 \caption{Fig.\ref{fig:fig1}. Temperature dependence of the zero-field
 {\it linear} thermal
 conductivity $\kappa _L(T,0)/\kappa _n$ for different
 values of the dimensionless parameter $\beta _L(0)$.}
  \label{fig:fig1}
\end{figure}

Despite a rather simplified nature of our model, it seems to quite
reasonably describe the behavior of the LTC for all temperatures.
Indeed, in the absence of an applied electric field ($E=0$), the LTC
is isotropic (as expected), $\kappa _{\alpha \beta}(T,0)=\delta
_{\alpha \beta}\kappa _L(T,0)$ where $\kappa _L(T,0)=\kappa _0[\eta
(T,0)+2\beta _L(T)\nu (T,0)]$ vanishes at zero temperature and
reaches a normal state value $\kappa _n\equiv \kappa _L(T_c,0)=(\pi
/2)\kappa _0$ at $T=T_c$. Figure~\ref{fig:fig1} shows the temperature
dependence of the normalized LTC $\kappa _L (T,0)/\kappa _n$ for
different values of the dimensionless parameter $\beta _L(0)=2\pi
I_c(0)L_{0}/\Phi _0$. As it is clearly seen, with increasing of this
parameter, the LTC evolves from a flat-like pattern (for a relatively
small values of $L_{0}$) to a low-temperature maximum (for higher
values of $\beta _L(0)$). Notice that the peak temperature $T_p$ is
practically insensitive to the variation of inductance parameter
$L_{0}$ while being at the same time strongly influenced by
resistivity $R_n$. Indeed, the presented here curves correspond to
the resistance ratio $r_n=R_0/R_n=1$ (a highly resistive state). It
can be shown that a different choice of $r_n$ leads to quite a
tangible shifting of the maximum. Namely, the smaller is the normal
resistance between grains $R_n$ (or the better is the quality of the
sample) the higher is the temperature at which the peak is developed.
As a matter of fact, the peak temperature $T_p$ is related to the
so-called phase-locking temperature $T_J$ (which marks the
establishment of phase coherence between the adjacent grains in the
array and always lies below a single grain superconducting
temperature $T_c$) which is usually defined via an average (per
grain) Josephson coupling energy as~\cite{13} $J(T_J,r_n)=k_BT_J$. In
particular, for $T\simeq T_c$, it can be shown analytically that
$T_J(r_n)$ indeed increases with $r_n$ as $T_J(r_n)/T_c \simeq
r_n/(1+r_n)$.

\begin{figure}
 \centerline{\includegraphics[width=90mm]{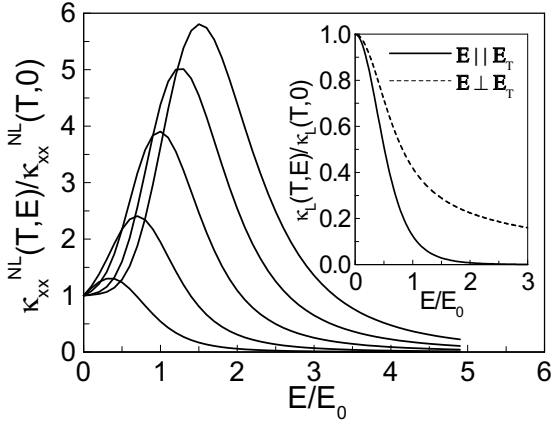}}
 \caption{Fig.\ref{fig:fig2}. Electric field dependence of the
 {\it nonlinear} thermal conductivity
  $\kappa _{xx}^{NL}(T,E)/\kappa _{xx}^{NL}(T,0)$ for different
  values of the applied thermal gradient $\epsilon _T=S_0|\nabla T|/E_0$
  ($\epsilon _T=0.2; 0.4; 0.6; 0.8; 1.0$, increasing from bottom to top).
 Inset: Electric field dependence of the {\it linear}
 thermal conductivity $\kappa _L(T,E)/\kappa _L(T,0)$ for parallel
 (${\bf E} \| \nabla T$) and perpendicular (${\bf E}
\bot \nabla T$) configurations.}
  \label{fig:fig2}
\end{figure}

{\it 3.2. Electric field effects.} Turning to the discussion of the
LTC behavior in applied electric field, let us demonstrate first of
all its anisotropic nature. For simplicity (but without losing
generality), we assume that ${\bf E}=(E,0,0)$ and $\nabla T=(\nabla
_xT, \nabla _yT,0)$. Such a choice of the external fields allows us
to consider both parallel $\kappa _{xx}(T,E)$ and perpendicular
$\kappa _{yy}(T,E)$ components of the LTC corresponding to the two
most interesting configurations, ${\bf E} \| \nabla T$ and ${\bf E}
\bot \nabla T$, respectively. Inset in Figure~\ref{fig:fig2}
demonstrates the predicted electric field dependence of the
normalized LTC $\kappa _L(T,E)/\kappa _L(T,0)$ for both
configurations taken at $T=0.2T_c$ (with $r_n=1$ and $\beta
_L(0)=1$). First of all, we note that both components of the LTC are
{\it decreasing} with increasing of the field $E/E_0$. And secondly,
the normal component $\kappa _{yy}$ decreases more slowly than the
parallel one $\kappa _{xx}$, suggesting thus some kind of anistropy
in the system. In view of the structure of Eq.(14), the same behavior
is also expected for the temperature dependence of the field-induced
LTC, that is $\Delta \kappa _L(T,E)/\kappa _L(T,0)<0$ for all fields
and temperatures. In terms of the absolute values, for $T=0.2T_c$ and
$E=E_0$, we obtain $[\Delta \kappa _L(T,E)/\kappa _L(T,0)]_{xx}=90\%$
and $[\Delta \kappa _L(T,E)/\kappa _L(T,0)]_{yy}=60\%$ for {\it
attenuation} of LTC in applied electric field.

{\bf 4. Nonlinear thermal conductivity: Giant field-induced effects.}
Let us turn now to the most intriguing part of this paper and
consider a {\it nonlinear} generalization of the Fourier law and very
unusual behavior of the resulting {\it nonlinear} thermal
conductivity (NLTC) under the influence of an applied electric field.
In what follows, by the NLTC we understand a $\nabla T$-dependent
thermal conductivity $\kappa _{\alpha \beta}^{NL}(T,{\bf E})\equiv
\kappa _{\alpha \beta }(T,{\bf E};\nabla T)$ which is defined as
follows
\begin{equation}
\kappa _{\alpha \beta}^{NL}(T,{\bf E})\equiv -\frac{1}{V}\left
[\frac{\partial \overline{<Q_{\alpha}>}}{\partial (\nabla
_{\beta}T)}\right ]_{\nabla T\neq 0}
\end{equation}
with $\overline{<Q_{\alpha}>}$ given by Eq.(11).

Repeating the same procedure as before, we obtain finally for the
relevant components of the NLTC tensor
\begin{gather}
\kappa _{\alpha \beta}^{NL}(T,{\bf E})=\kappa _0[\delta _{\alpha
\beta}\eta (T,\epsilon _{eff})\\ \notag
+\beta _L(T)\nu (T,\epsilon _{eff})D_{\alpha \beta}(\epsilon
_{eff})]\notag,
\end{gather}
where
\begin{equation}
D_{\alpha \beta}(\epsilon _{eff})=f_{\alpha \beta}(\epsilon _{eff})+
\epsilon _{T}^{\gamma}g_{\alpha \beta \gamma} (\epsilon _{eff})
\end{equation}
with
\begin{gather}
g_{\alpha \beta \gamma} (\epsilon )=\frac{1}{8}[(\delta _{\alpha
\beta}\epsilon _{\gamma}+ \delta _{\alpha \gamma}\epsilon _{\beta} +
\delta _{\gamma \beta}\epsilon _{\alpha})B(\epsilon )\\ \notag
 + 3\epsilon
_{\alpha} \epsilon _{\beta} \epsilon _{\gamma} C(\epsilon )]\notag
\end{gather}
and
\begin{equation}
C(\epsilon)=\frac{3+11\epsilon ^2-11\epsilon ^4-3\epsilon
^6}{\epsilon ^4(1+\epsilon ^2)^4}-\frac{3}{\epsilon ^5}\tan
^{-1}\epsilon
\end{equation}
Here, $\epsilon _{eff}^{\alpha}=\epsilon ^{\alpha} -\epsilon
_T^{\alpha}$ where $\epsilon _{\alpha}=E_{\alpha}/E_0$ and $\epsilon
_T^{\alpha}=E_T^{\alpha}/E_0$ with $E_T^{\alpha}=S_0\nabla
_{\alpha}T$; other field-dependent parameters ($\eta$, $\nu$, $B$ and
$f_{\alpha \beta}$) are the same as before but with $\epsilon \to
\epsilon _{eff}$.

As expected, in the limit $E_T \to 0$ (or when $E\gg E_T$), from
Eq.(22) we recover all the results obtained in the previous section
for the LTC. Let us see now what happens when the "intrinsic"
thermoelectric field ${\bf E_T}=S_0\nabla T$ becomes comparable with
an applied electric field ${\bf E}$. Figure~\ref{fig:fig2} (main
frame) depicts the resulting electric field dependence of the
parallel component of the NLTC tensor $\kappa _{xx}^{NL}(T,E)$ for
different values of the dimensionless parameter $\epsilon _T=E_T/E_0$
(the other parameters are the same as before). As it is clearly seen
from this picture, in a sharp contrast with the field behavior of the
previously considered {\it linear} TC, its {\it nonlinear} analog
evolves with the field quite differently. Namely, NLTC strongly {\it
increases} for small electric fields ($E<E_m$), reaches a pronounced
maximum at $E=E_m=\frac{3}{2}E_T$, and eventually declines at higher
fields ($E>E_m$). Furthermore, as it directly follows from the very
structure of Eq.(22), a similar "reentrant-like" behavior of the {\it
nonlinear} thermal conductivity will occur in its temperature
dependence as well. Even more remarkable is the absolute value of the
field-induced enhancement. According to Figure~\ref{fig:fig2} (main
frame), it is easy to estimate that near maximum (with $E=E_m$ and
$E_T=E_0$) and for $T=0.2T_c$,  one gets $\Delta \kappa
_{xx}^{NL}(T,E)/\kappa _{xx}^{NL}(T,0)\simeq 500\%$.

{\bf 5. Discussion.} To understand the above-obtained rather unusual
results, let us take a closer look at the field-induced behavior of
the Josephson voltage in our system (see Eq.(7)). Clearly, strong
heat conduction requires establishment of a quasi-stationary (that is
nearly zero-voltage) regime within the array. In other words, the
maximum of the thermal conductivity in applied electric field should
correlate with a minimum of the total voltage in the system,
$V(E)\equiv (\frac{\hbar}{2e})<\frac{\partial \phi _{ij}(t)}{\partial
t}>=V_0(\epsilon -\epsilon _T)$ where $\epsilon \equiv E/E_0$ and
$V_0=E_0d=\hbar /2e\tau $ is a characteristic voltage. For linear TC
(which is valid only for small thermal gradients with $\epsilon
_T\equiv E_T/E_0\ll 1 $), the average voltage through an array
$V_L(E)\simeq V_0(E/E_0)$ has a minimum at zero applied field (where
LTC indeed has its maximum value, see the inset of
Figure~\ref{fig:fig2}) while for nonlinear TC (with $\epsilon
_T\simeq 1 $) we have to consider the total voltage $V(E)$ which
becomes minimal at $E=E_T$ (in a good agreement with the predictions
for NLTC maximum which appears at $E=\frac{3}{2}E_T$, see the main
frame of Figure~\ref{fig:fig2}).

To complete our study, let us estimate an order of magnitude of the
main model parameters. Starting with applied electric fields $E$
needed to observe the above-predicted nonlinear field effects in
granular superconductors, we notice that according to
Figure~\ref{fig:fig2}, the most interesting behavior of NLTC takes
place for $E\simeq E_0$. Taking $d\simeq 10\mu m$ and $\tau \simeq
10^{-9}s$ for typical values of the average grain size and
characteristic Josephson tunneling time (valid for conventional
JJs~\cite{14} and HTS ceramics~\cite{10}), we get $E_0=\hbar
/(2ed\tau )\simeq 2\times 10^{-2}V/m$ for the characteristic electric
field (which is surprisingly lower than the typical fields needed to
observe a critical current enhancement in HTS ceramics~\cite{4,5}).
On the other hand, the maximum of NLTC occurs when this field nearly
perfectly matches an "intrinsic" thermoelectric field $E_T=S_0|\nabla
T|$ induced by an applied thermal gradient, that is when $E\simeq
E_0\simeq E_T$. Using $S_0 \simeq 0.5\mu V/K$ for the zero-field
value of the {\it linear} Seebeck coefficient~\cite{10,14}, we obtain
$|\nabla T|_E\simeq E_0/S_0\simeq 4\times 10^4K/m$ for the
characteristic value of an applied thermal gradient. Finally, taking
as an example granular aluminum films with phonon dominated heat
transport~\cite{15} (with $\kappa _{ph}(T)\simeq 2\times 10^{-7}W/mK$
at $T=T_J\simeq 0.2T_c$), let us estimate the absolute value of the
predicted here zero-field electronic contribution $\kappa _e(T)\equiv
\kappa _L(T,0)$ at $T=0.2T_c$. Recalling that within our model the
scattering of normal electrons is due to the presence of mutual
inductance between the adjacent grains $L_0$, and assuming
that~\cite{13} $L_0\simeq \mu _0d\simeq 4\pi \times 10^{-12}H$ and
$V\simeq Nd^2l$ with $l\simeq 0.5mm$ ($l$ is a film's thickness), we
obtain $\kappa _e(T=0.2T_c)\simeq \beta _L(0)\times 10^{-7}W/mK$ for
a rough estimate of the electronic contribution  to the discussed
here inductance-driven effect. Correspondingly, we get $\kappa
_e(0.2T_c)/\kappa _{ph}(0.2T_c)\simeq \beta _L(0)/2$ for the ratio,
where $\beta _L(0)=2\pi I_c(0)L_{0}/\Phi _0$ (for example, $\beta
_L(0)\simeq 4$ for $I_c(0)=10^{-4}A$). Thus, depending mainly on the
value of the critical current $I_c(0)$ and mutual inductance between
adjacent grains $L_0$, the thermal conductivity of specially prepared
granular alumina films will be dominated by either phonon (for small
$\beta _L(0)$) or electronic (for large $\beta _L(0)$) contribution.
Undoubtedly, the above estimates suggest quite a realistic
possibility to observe the predicted non-trivial behavior of the
thermal conductivity in granular superconductors and artificially
prepared Josephson junction arrays. We hope that the presented here
results will motivate further theoretical and experimental studies of
this interesting problem.


\begin{thebibliography}{99}
\bibitem{1} {\em Proceedings of the Conference "Mesoscopic and Strongly
Correlated Electron Systems", Chernogolovka, 1997}, Ed. by V.F.
Gantmakher and M.V. Feigel'man, Phys.Usp. {\bf 41} (2) (1998); {\em
Mesoscopic and Strongly Correlated Electron Systems-II}, Ed. by M.V.
Feigel'man, V.V. Ryazanov and V.B. Timofeev, Phys. Usp. (Suppl.) {\bf
44} (10) (2001).
\bibitem{2} See, e.g., a special issue on JJAs in
{\em Studies of High Temperature Superconductors}, vol. {\bf 39}, Ed.
by A. Narlikar and F. Araujo-Moreira (Nova Science, New York, 2001).
\bibitem{3} L.B. Ioffe, M.V. Feigel'man, A. Ioselevich et al., Nature
{\bf 415}, 503 (2002).
\bibitem{4} See, e.g., T.S. Orlova and B.I. Smirnov,
Supercond. Sci. Technol. {\bf 7}, 899 (1994); T.S. Orlova, B.I.
Smirnov and J.Y. Laval, Phys. Solid State {\bf 43}, 1007 (2001).
\bibitem{5} T.S. Orlova, B.I. Smirnov, J.Y. Laval et al.,
Supercond. Sci. Technol. {\bf 12}, 356 (1999).
\bibitem{6} A.L. Rakhmanov and A.V. Rozhkov, Physica C
{\bf 267}, 233 (1996).
\bibitem{7} D. Dominguez, C. Wiecko and J.V. Jos\'e,
Phys. Rev. Lett. {\bf 83}, 4164 (1999).
\bibitem{8}  J. Mannhart, Supercond. Sci. Technol.
{\bf 9}, 49 (1996).
\bibitem{9}  S.A. Sergeenkov and J.V. Jos\'e, Europhys. Lett.
{\bf 43}, 469 (1998).
\bibitem{10} S.A. Sergeenkov, JETP Lett. {\bf 67}, 680 (1998).
\bibitem{11} A.-L. Eichenberger, J. Affolter, M. Willemin et al.,
 Phys. Rev. Lett. {\bf 77}, 3905 (1996).
\bibitem{12} R. Meservey and B.B. Schwartz, in
{\it Superconductivity}, vol.1, ed. by R.D. Parks (M. Dekker, NY,
1969), p.117.
\bibitem{13} L. Leylekian, M. Ocio, L.A. Gurevich et al.,
JETP {\bf 85}, 1138 (1997).
\bibitem{14} G.I. Panaitov, V.V. Ryazanov, A.V. Ustinov et al.,
Phys. Lett. A {\bf 100}, 301 (1984).
\bibitem{15} J. Deppe and J.L. Feldman, Phys. Rev. B {\bf 50}, 6479 (1994).
\end{thebibliography}
\end{document}